\documentclass[reprint, amsmath,amssymb,superscriptaddress,aps]{revtex4-1}
\usepackage{latexsym}
\usepackage{amsfonts}
\usepackage{mathrsfs}
\usepackage{amssymb,amscd}
\usepackage[all]{xy}
\usepackage{amsmath}
\usepackage{amsthm}
\usepackage{fancyhdr}
\usepackage{fancybox}
\usepackage{xcolor}
\usepackage{hyperref}
\usepackage{verbatim}
\usepackage{makeidx}
\usepackage{indentfirst}
\usepackage[english]{babel}
\usepackage{graphicx}
\usepackage{listings}
\usepackage{subfigure}
\usepackage{multirow}
\usepackage{physics}
\usepackage{amsmath}
\usepackage{framed}
\usepackage{bbold}
\usepackage{cases}
\usepackage{dcolumn}
\usepackage{bm}
\usepackage{siunitx}
\begin{document}

\title{Ferromagnetic resonance linewidth in coupled layers with easy-plane and perpendicular magnetic anisotropies}

\author{Jun-Wen Xu}
\affiliation{Center for Quantum Phenomena, Department of Physics, New York University, New York, NY 10003, USA}
\author{Volker Sluka}
\email{vs1568@nyu.edu}
\affiliation{Center for Quantum Phenomena, Department of Physics, New York University, New York, NY 10003, USA}
\author{Bartek Kardasz}
\affiliation{Spin Transfer Technologies, Inc., Fremont, California 94538, USA}
\author{Mustafa Pinarbasi}
\affiliation{Spin Transfer Technologies, Inc., Fremont, California 94538, USA}
\author{Andrew D. Kent}
\email{andy.kent@nyu.edu}
\affiliation{Center for Quantum Phenomena, Department of Physics, New York University, New York, NY 10003, USA}

\date{\today}

\begin{abstract}
Magnetic bilayers with different magnetic anisotropy directions are interesting for spintronic applications as they offer the possibility to engineer tilted remnant magnetization states. 
We investigate the ferromagnetic resonance (FMR) linewidth of modes associated with two interlayer exchange-coupled ferromagnetic layers, the first a CoNi multilayer with a perpendicular magnetic anisotropy, and the second a CoFeB layer with an easy-plane anisotropy.
For antiferromagnetic interlayer exchange coupling, elevated FMR linewidths are observed below a characteristic field. This is in contrast to what is found in uncoupled, ferromagnetically coupled and single ferromagnetic layers in which the FMR linewidth increases monotonically with field.
We show that the characteristic field at which there is a dramatic increase in FMR linewidth can be understood using a macrospin model with Heisenberg-type exchange coupling between the layers.
\end{abstract}

\pacs{Valid PACS appear here}
\maketitle

\section{Introduction}
Magnetic multilayers are widely used in spintronics devices for information processing and storage~\cite{Fert}.
An essential means for engineering the coupling between two ferromagnetic (FM) layers is by interlayer exchange coupling (IEC)
mediated by an intervening nonmagnetic (NM) layer. The sign and strength of the IEC changes with the thickness of the NM; typically
the coupling oscillates between FM and AFM while the coupling strength decreases with increasing interlayer thickness~\cite{37.coupling}.

One device concept of interest at present is spin transfer torque (STT) magnetic random access memory (MRAM)~\cite{KentWorledge2015}.
It offers a high density, non-volatile, short switching time and low power consumption information storage method. The basic
structure of the memory cell consists of a magnetic tunnel junction, two FM layers separated by a thin insulating barrier layer.
In the conventional layer stacks used in MRAM design, the stable magnetic configurations exhibit collinear magnetization alignment. 
While these alignments provide maximum contrast in terms of magnetoresistance, and thus optimal read-out conditions, such alignment 
has disadvantages from the perspective of STT, since the torque exerted is proportional to the cross product of the magnetization vectors~\cite{STT}.
In other words, small initial deviations from perfect alignment are needed to initiate STT-induced switching. Such deviations are present due to thermal 
fluctuations, however, the resulting initial torques are small.  For this reason, there is interest in devices with non-collinear magnetic configurations.
Furthermore, it is important to have a better understanding of the magnetic relaxation of the ferromagnetic resonance (FMR) modes in such structures, 
since the time scale of relaxation determines the switching speed of the system. Spin relaxation is described by the phenomenological Gilbert damping
and FMR is a convenient way to measure the damping directly from the spectrum~\cite{1.alpha,6.FMR}.

\section{Experimental setup}
In order to investigate magnetization dynamics in a system with non-collinear magnetic configurations, we have conducted experiments on samples 
containing two FM layers exhibiting different forms of the magnetic anisotropy: one with easy-plane anisotropy, the other one with perpendicular magnetic 
anisotropy (PMA). As the PMA material we use a CoNi multilayer, while a $\text{Co}_{60} \text{Fe}_{20} \text{B}_{20}$ (CoFeB) film serves as the easy-plane
magnetic layer. Both layers are coupled via IEC by a Ru interlayer: Using dc magnetron sputtering, 
[Ta(5)/Cu(3)/Ni(0.65)/Co(0.3)/ [Ni(0.6)/Co(0.2)]$_5$/Co(0.18)/Ru($t_\text{Ru}$)/CoFeB(3)/Ta(3)] samples were deposited on $\SI{150}{mm}$ oxidized silicon wafer.
In order to investigate different strengths and signs of coupling between the CoNi and CoFeB layers, the thickness of Ru ($t_\text{Ru}$), was varied along one
coordinate on the wafer in a wedge-like manner, $t_\text{Ru}$ ranges from 0.71 to $\SI{1.17}{nm}$ across the wafer. For the FMR experiment, we cut
$2 \times \SI{2}{mm^2}$-sized pieces from the wafer, where the lateral distance between pieces is $8\,\mathrm{mm}$. Therefore, to a good approximation, 
we can treat the samples as having uniform Ru thickness within each piece. Besides these coupled-layer samples, we also studied two samples with a single 
CoNi FM layer [Ta(5)/Cu(3)/Ni(0.65)/Co(0.3)/[Ni(0.6)/Co(0.2)]$_5$/ Co(0.18)/Ru(3)] and a single CoFeB FM layer [Ta(5)/ Cu(3)/Ru(1)/CoFeB(3)/Ta(3)], as reference. 

\begin{figure}[ht!]
    \centering
    \includegraphics[width=0.47\textwidth]{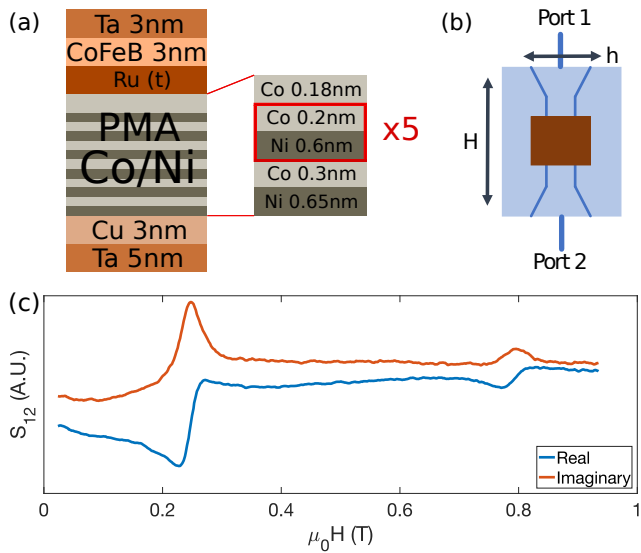}
    \caption{(a) Multilayer sample, with varying Ru thickness. (b) Schematic of FMR measurement using a two-port VNA. $H$ is the dc applied magnetic field and $h$ is ac Oersted field. (c) Example normalized real and imaginary S-parameter data versus applied field for a sample with $t_\text{Ru}= 0.99$ nm at $\SI{17}{GHz}$ ac frequency.}
    \label{fig:multilayer}
\end{figure}

In order to measure the FMR spectra, we apply a dc magnetic field parallel to the sample plane.
The sample is mounted on a waveguide; an ac current through the waveguide provides a small oscillating Oersted field 
perpendicular to the dc applied field and drives the magnetization into a small angle precession, which has its maximum amplitude when the resonance condition is satisfied. 
We use a vector network analyzer (VNA) to measure the magnetization dynamics by determining the effective field- and frequency-dependent load the sample adds to the waveguide.
Fixing the frequency of the ac field, the dc magnetic field $\mu_{0} H$ is swept from 0 to \SI{1}{T}.
We collect the transmission ($S_{1 2}$) and reflection ($S_{1 1}$ and $S_{2 2}$) spectra with a VNA~\cite{8.S,9.S}.
For each spectrum, we extract data from the S-parameter providing the highest signal to noise ratio.
By fitting both real and imaginary parts of the S-parameter with a linear combination of symmetric and antisymmetric Lorentzian functions, we obtain the peak position and full width at half maximum (FWHM), which correspond to the resonance field and linewidth, respectively. This analysis is done for the real and imaginary parts of the chosen S parameter, and the resulting values for peak position and FWHM are averaged and plotted as a function of the applied field (Fig. \ref{fig:line}).

\section{Experimental results}
The multilayer samples used in the measurement are taken from the same wafer used in previous research~\cite{0.Lorenzo}. 
Based on the previous results, the samples with a Ru thickness of $\SI{1.02}{nm}$ and smaller exhibit AFM coupling between the layers while for the samples with Ru thickness $\SI{1.05}{nm}$ and larger, the IEC is ferromagnetic.
In order to investigate how the coupling affects the FMR linewidth, samples with Ru thicknesses of 0.95, 0.99, 1.02, 1.05 and $\SI{1.09}{nm}$ were studied.
Their coupling strength $J$ is shown in Table \ref{tb:J}.

\begin{table}[t]
    \centering
    \begin{tabular}{l|r r r r r }
        \hline
        \hline
        $t_\text{Ru}$ ($\SI{}{nm}$)     & 0.95   & 0.99   & 1.02   & 1.05  & 1.09  \\ 
        $J$ ($\SI{}{mJ/m^2}$)           & -0.330 & -0.270 & -0.116 & 0.249 & 0.365 \\ 
        $\mu_0 H_{k 1}$ ($\SI{}{T}$)          & 0.186  & 0.179  & 0.142  & 0.204 & 0.181 \\ 
        $\mu_0 H_{k 2}$ ($\SI{}{T}$)          & -0.904 & -0.911 & -0.948 & -0.886& -0.909\\ 
        \hline
        \hline
    \end{tabular}
    \caption{Sample parameters determined in Ref.~\cite{0.Lorenzo}. $t_\text{Ru}$ is the Ru layer thickness and $J$ is the IEC. 
    $\mu_0 H_{k 1}$ and   $\mu_0 H_{k 2}$ are the CoNi and CoFeB effective perpendicular magnetic anisotropy fields, respectively.}
    \label{tb:J}
\end{table}

\begin{figure}[b]
    \centering
    \includegraphics[width=0.48\textwidth]{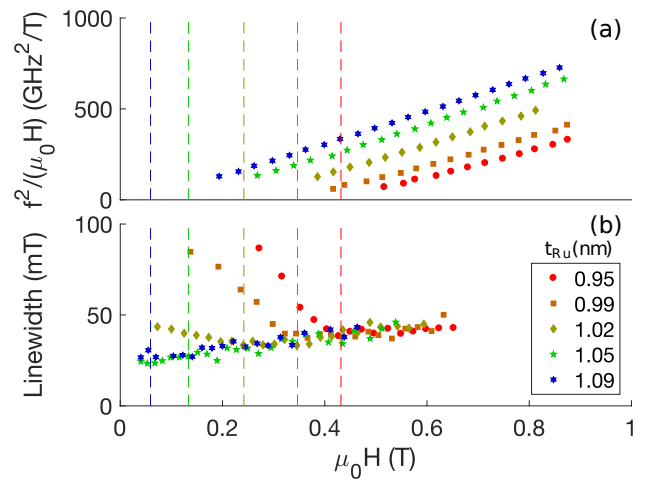}
    \caption{In-plane FMR measurement of the samples listed in Table~\ref{tb:J}. (a) Resonance position of the CoNi mode for different interlayer thicknesses. The dashed lines indicate the intercept of a linear fit to the data with the x-axis, which represents the in-plane saturation field. With increasing IEC, the saturation field decreases. (b) Linewidth of the CoFeB mode. When the layers are FM coupled, the linewidth decreases monotonically as the field decreases. However, when the layers are AFM coupled, the linewidth first decreases with decreasing field but then increases when the applied field is less than the saturation field.}
    \label{fig:line}
\end{figure}

Due to the presence of two FM layers, there exist two FMR modes. 
In the limit of vanishing IEC, these modes will become independent FMR modes of the CoNi and CoFeB layers. At high in-plane applied fields deeply in the saturated regime, the CoFeB mode will have a significantly higher frequency than the CoNi mode. This hierarchy is maintained for non-zero IEC, when the two modes hybridize. Therefore, for simplicity,  throughout this paper, we will refer to the two modes as the CoNi and the CoFeB modes, even in the case of nonzero coupling.

The IEC leads to contributions to the effective field acting on each layer. 
For high enough applied fields, when the layers are saturated, the IEC therefore leads to a shift in the resonance fields.
Figure \ref{fig:line}(a) shows the relation between applied field and resonance frequency of the CoNi mode.
Since CoNi has perpendicular anisotropy, when the applied field is small its magnetization is not saturated in the applied field direction.
The saturation fields correspond to the intercepts of the data sets shown in Fig.~\ref{fig:line}(a) with the horizontal axis.
The shift of the saturation field for different samples can be understood in the following way: due to the IEC the two magnetic layers experience a coupling field from each other.
Considering the saturated case, for FM coupling, layer $i$ experiences an exchange coupling field in the same direction as the magnetization of layer $j$, while in the case of AFM coupling, the coupling field points in the opposite direction with respect to the magnetization of layer $j$. Consistent with this picture, Fig.~\ref{fig:line}(a) shows an increasing saturation field when the coupling becomes more AFM. 

Figure \ref{fig:line}(b) displays the linewidth of the CoFeB mode as a function of the applied field.
In the case of FM coupling, the linewidth increases with increasing frequency and increasing resonance field, which is similar to what would be observed in a single easy-plane layer in in-plane field~\cite{4.Kittel}.
However, the figure also shows that in AFM coupled cases, when going from high to low fields, below a certain sample-dependent characteristic field, the linewidth increases significantly. This behavior is very different from what is typically observed for single layers.

\section{Modeling and simulation}
The net magnetic field acting on a uniformly magnetized layer, known as the effective field, is the sum of applied field, anisotropy field and coupling field.
When the applied field is smaller than the saturation field, the magnetization of the CoNi layer is not aligned with the field (i.e. an in-plane) direction.
Decreasing the applied field further will increase the out-of-plane angle between magnetization of CoNi layer and the applied field.
Due to the AFM coupling, this increasing angle {\em increases} the coupling field in the applied field direction acting on the CoFeB layer. 
Thus, it slows down the decrease of the effective field experienced by the CoFeB layer; the CoNi layer can partially screen the applied field experienced by the CoFeB layer.
In the experiment, we fix the frequency and sweep the applied field to determine the linewidth. 
When the applied field is smaller than the saturation field, due to this screening effect, a change in the applied field results in a corresponding smaller change in the CoFeB effective field and thus an increase in the FMR linewidth.

In order to model the experimental results, we assume a Heisenberg-type exchange coupling between the layers, with film areal energy density \cite{0.Lorenzo,11.demagnetization,12.demagnetization,13.demagnetization}
\begin{equation}
    \begin{split}
        \sigma_E(\vec{m}_1, \vec{m}_2) =& - \mu_0 M_{s 1} d_1 \vec{H} \cdot \vec{m}_1 - \frac{1}{2} \mu_0 M_{s 1} d_1 H_{k 1} m'^2_{z 1}\\
        & - \mu_0 M_{s 2} d_2 \vec{H} \cdot \vec{m}_2 - \frac{1}{2} \mu_0 M_{s 2} d_2 H_{k 2} m'^2_{z 2}\\
        &  - J \vec{m}_1 \cdot \vec{m}_2,
    \end{split}
\end{equation}
where $\vec{m}_i = m'_{x i} \hat{x}' + m'_{y i} \hat{y}' + m'_{z i} \hat{z}'$ is a unit vector in the magnetization direction of each layer in the lab frame, with $i = 1$ representing the CoNi layer and $i = 2$ representing the CoFeB layer.
Here, we denote our lab frame as $\hat{x}'$, $\hat{y}'$, $\hat{z}'$, where $\hat{x}'$ is aligned with the applied field and $\hat{z}'$ is perpendicular to the layer plane. The
1st and 3rd term, which are proportional to the applied field $\vec{H}$, represent the Zeeman energy; the 2nd and 4th term, which are proportional to $m'^2_{z i}$, represent the magnetic anisotropy energy; the last term is the coupling energy.
The effective anisotropy is $H_{k i} \equiv H_{k i}^{(0)} - M_{s i}$, where $H_{k i}^{(0)}$ is the intrinsic anisotropy field and $M_{s i}$ is the saturation magnetization.
For the CoNi layer, with perpendicular easy axis, $H_{k 1}^{(0)} > 0$ and for CoFeB layer, with in-plane easy axis, $H_{k 2}^{(0)} < 0$.
$J$ is the coupling strength; $J > 0$ is FM coupling and $J < 0$ AFM coupling.
$d_i$ is the thickness of each magnetic layer and $\mu_0$ is the vacuum permeability.

\begin{figure}[t]
    \centering
    \includegraphics[width=0.48\textwidth]{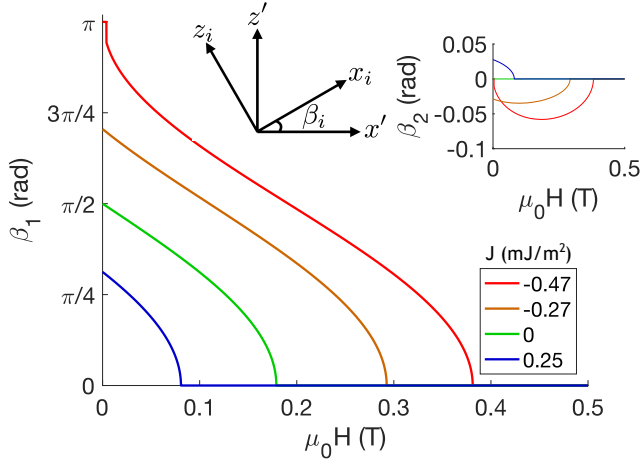}
    \caption{The CoNi magnetization angle  $\beta_1$ (main figure) and CoFeB magnetization angle $\beta_2$ (right inset) obtained by minimizing the energy as a function of applied field. The greater the AFM coupling strength, the larger field required to saturate the CoNi layer. The CoFeB magnetization has a  small out of plane component associated with the small ratio of the IEC field to the in-plane anisotropy field. Middle inset: schematic showing the $\beta$ angles.}
    \label{fig:angle}
\end{figure}

When the applied field is small, the magnetization of each layer may not be saturated and the precession axis need not be aligned with $\hat{x}'$. 
Thus, we rotate our lab frame by $\beta_i$ along $\hat{y}'$, to $\hat{x}_i$, $\hat{y}_i$, $\hat{z}_i$, to the so-called ground state frame, where $\hat{x}_i$ is the precession axis of each layer. 
When minimizing $\sigma_E$, we determine $\beta_i$ numerically. Figure~\ref{fig:angle}  shows a numerical solution for $\beta_i$. 
Since for the CoFeB layer, the coupling strength is smaller compared with the anisotropy field, $|\beta_2| < |\beta_1|$ in general.
The smaller the coupling strength, the larger the saturation field.
When the coupling becomes AFM, the CoNi layer provides a coupling field on the CoFeB layer opposite the direction of the CoNi magnetization, which thus requires a larger applied field to compensate.

The effective field in the lab frame is:
\begin{equation}
    \vec{H}_i(m'_{x i}, m'_{y i}, m'_{z i}) = - \frac{1}{\mu_0 M_{s i} d_i} \grad_{\vec{m}'_i} \sigma_E.
\end{equation}
After numerically determining $\beta_i$ at a given applied field, we rewrite $\vec{m}'_i$ in ground state frame, $\vec{m}_i = m_{x i} \hat{x}_i + m_{y i} \hat{y}_i + m_{z i} \hat{z}_i$ by using the rotation matrix shown in the appendix.
The magnetization precesses around the $\hat{x}_i$ axis. 
Considering a small angle precession, $m_{x i} \approx 1$, $m_{y i}$ and $m_{z i}$ can be expanded around 0. 
On the other hand, we can rewrite the effective field in ground state frame, $\vec{H}_i = \vec{H}_i(m_{x i}, m_{y i}, m_{z i})$.

Finally, we use LLG equation in the ground state frame to find the equation of motion of the magnetization of both layers~\cite{15.Gilbert}:
\begin{equation}
    \dv{\vec{m}_i}{t} = - \gamma_i (\vec{m}_i \cross \vec{H}_i) + \alpha_i \left( \vec{m}_i \cross \dv{\vec{m}_i}{t}  \right) + h \cos(\omega t) \hat{y}_i,
\label{Eq:LLG}    
\end{equation}
where, on the right hand side, the first term is the precession term, the second term is damping term and the third term is oscillating field.
$\alpha_i$ and $\gamma_i$ represent damping coefficient and gyromagnetic ratio, respectively.
By moving 2nd term to the left hand side and matrix inversion, we can rewrite Eqn.~\ref{Eq:LLG} as
\begin{equation}
    \dv{m}{t} = \Gamma \Lambda (\Xi m + C),\label{eq:FT}
\end{equation}
where
\begin{equation}
    m(t) \equiv \begin{pmatrix} m_{y1}(t)& m_{z1}(t)& m_{y2}(t)& m_{z2}(t) \end{pmatrix}^T,
\end{equation}
\begin{equation}
    \Xi = \Xi_\text{a} + \Xi_\text{c} + \Xi_\text{an}
\end{equation}
\begin{equation}
    C \equiv \begin{pmatrix} h \cos(\omega_0 t)& 0& h \cos(\omega_0 t)& 0 \end{pmatrix}^T.
\end{equation}
The modified gyromagnetic ratio matrix $\Gamma$, the Gilbert damping coefficient matrix $\Lambda$ and the elements of effective field matrices $\Xi_\text{a}$, $\Xi_\text{c}$, $\Xi_\text{an}$ are shown in the appendix.

In order to solve the coupled ODEs, we take the Fourier transformation of Eqn.~\ref{eq:FT}, $\tilde{f}(\omega) \equiv \frac{1}{\sqrt{2 \pi}} \int_{-\infty}^\infty f(t) e^{- i \omega t} \, \mathrm{d} t$ to get:\begin{equation}
    \tilde{m} = - \frac{\Gamma \Lambda \tilde{C}}{(\Gamma \Lambda \Xi - i \omega \mathbb{1})},
\end{equation}
where the amplitude of the magnetization is dominated by $1/|\det(\Gamma \Lambda \Xi - i \omega \mathbb{1})|$.

\begin{figure}[t]
    \centering
    \includegraphics[width=0.5\textwidth]{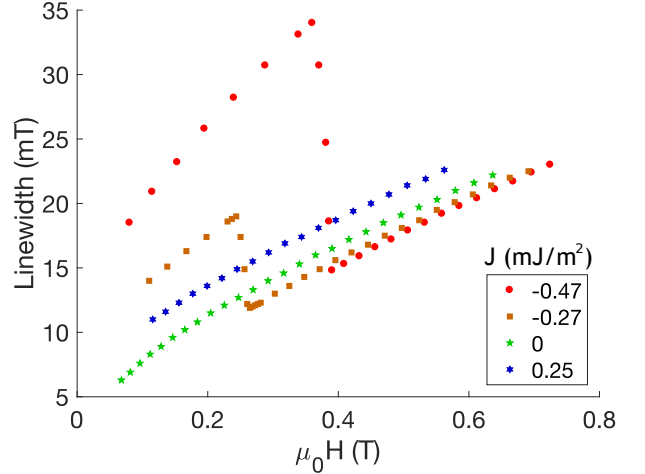}
    \caption{Computed CoFeB linewidth versus applied field. When the IEC is FM, the linewidth decreases with decreasing field. When the IEC is AFM, there is an abrupt increase in the linewidth when the field decreases below the saturation field.} 
    \label{fig:narrow}
\end{figure}

In the numerical simulation, we take $\alpha_1 = 0.0108$ and $\alpha_2 = 0.0053$, which come from the single layer sample measurements. 
We take $\gamma_1$ and $\gamma_2$ to be the electron's gyromagnetic ratio. We further
take $\mu_0 H_{k 1}^{(0)} = \SI{0.779}{T}$, $\mu_0 H_{k 2}^{(0)} = \SI{-0.128}{T}$, $\mu_0 M_{s 1} = \SI{0.600}{T}$ and $\mu_0 M_{s 2} = \SI{1.69}{T}$, to characterize the magnetic materials~\cite{0.Lorenzo}. Since our goal is to illustrate the mechanism leading to the observed increase in linewidth, rather than a quantitiative fit, we keep these values fixed and only vary the coupling strength $J$.

Figure~\ref{fig:narrow} shows the numerical solution obtained for different coupling strengths $J$, which matches our experimental results to a certain degree.
When the coupling strength is ferromagnetic, the linewidths increase monotonically with the applied field, which matches our experimental results.
In the case of AFM coupling, when sweeping the field from high to low values, the linewidths decrease with decreasing applied field, at least initially. 
But when the applied field reaches the saturation field, there is a strong enhancement in the linewidth.
This qualitatively reproduces the behavior seen in the experiments.
The saturation fields shown in Fig. \ref{fig:narrow} are consistent with the previous simulation results shown in Fig.~\ref{fig:angle}(a). 
The saturation field increases when the samples become more AFM coupled.
It shows that the basic cause of the increasing linewidth is the screening of the effective field associated with the AFM coupling.
However, when further decreasing the applied field, the linewidths also decrease, which is not observed experimentally. The model also does not explain the finite width
of the field region, over which linewidth increase takes place. A possible reason is that our model assumes two macrospins, which means we consider the magnetization to be 
homogeneous in the sample plane. When the field is small, the magnetization may form domains, which violates the macrospin assumption. In addition, two-magnon 
scattering may contribute to the observed linewidth, as well as inhomogeneous local fields arising from defects.

\section{Summary}
In this work we investigated the linewidth in an interlayer exchange-coupled bilayer system with mixed anisotropies. 
In samples with AFM coupling, below the saturation field, we observed a significant increase of the linewidth in the FMR spectra.
Analysis in the framework of a previously established macrospin model points towards a mechanism that can be regarded as an applied field screening effect that results from the combined action of the IEC and the CoNi magnetization tilting due to its PMA.  The reported screening effect could be exploited in future spin torque oscillator applications where it could help stabilize the oscillator frequency with respect to changes in the ambient magnetic field.

\section{Acknowledgements}
This research was supported by Spin Transfer Technologies, Inc. and in part by National Science Foundation under Grant No. DMR-1610416.

\medskip

\onecolumngrid
\section{appendix}
In this appendix we provide the matrices used in the main text. The rotation matrix to transform between the lab $\vec{m}'$ and ground state frame $\vec{m}$ is:
\begin{equation}
    \begin{pmatrix} m_{x i}\\ m_{y i}\\ m_{z i} \end{pmatrix} = \begin{pmatrix} \cos(\beta_i)& 0& -\sin(\beta_i)\\ 0& 1& 0\\ \sin(\beta_i)& 0& \cos(\beta_i) \end{pmatrix} \begin{pmatrix} m'_{x i}\\ m'_{y i}\\ m'_{z i} \end{pmatrix}.
\end{equation}
The following are the matrices used to solve the LLG equation Eqn.~\ref{eq:FT}. The modified gyromagnetic ratio matrix:
\begin{equation}
    \Gamma \equiv \begin{pmatrix} \frac{\gamma_1}{1 + \alpha_1^2}&&&\\& \frac{\gamma_1}{1 + \alpha_1^2}&&\\&& \frac{\gamma_2}{1 + \alpha_2^2}&\\&&& \frac{\gamma_2}{1 + \alpha_2^2} \end{pmatrix}.
\end{equation}
The Gilbert damping coefficient matrix is:
\begin{equation}
    \Lambda \equiv \begin{pmatrix} 1& -\alpha_1&&\\ \alpha_1& 1&&\\&& 1& -\alpha_2\\&& \alpha_2& 1 \end{pmatrix}.
\end{equation}
Finally, the effective field matrices are:
\begin{equation}
    \Xi_\text{a} \equiv \begin{pmatrix} & - H \cos(\beta_1)&&\\ H \cos(\beta_1)&&&\\&&& - H \cos(\beta_2)\\&& H \cos(\beta_2)& \end{pmatrix},
\end{equation}
\begin{equation}
    \Xi_\text{c} \equiv \begin{pmatrix} & -\frac{J}{\mu_0 M_{s1} d_1} \cos(\beta_1 - \beta_2)&& \frac{J}{\mu_0 M_{s1} d_1} \cos(\beta_1 - \beta_2)\\ \frac{J}{\mu_0 M_{s1} d_1} \cos(\beta_1 - \beta_2)&& -\frac{J}{\mu_0 M_{s1} d_1}&\\&  \frac{J}{\mu_0 M_{s2} d_2} \cos(\beta_1 - \beta_2)&& - \frac{J}{\mu_0 M_{s2} d_2} \cos(\beta_1 - \beta_2)\\ - \frac{J}{\mu_0 M_{s2} d_2}&& \frac{J}{\mu_0 M_{s2} d_2} \cos(\beta_1 - \beta_2)& \end{pmatrix},
\end{equation}
\begin{equation}
    \Xi_\text{an} \equiv \begin{pmatrix} & H_{k1} \cos(2\beta_1)&&\\ H_{k1} \sin[2](\beta_1)&&&\\&&& H_{k2} \cos(2\beta_2)\\&& H_{k2} \sin[2](\beta_2)& \end{pmatrix}.
\end{equation}

\twocolumngrid

\end{document}